\documentclass[graybox]{svmult}

\usepackage{mathptmx}       
\usepackage{helvet}         
\usepackage{courier}        
\usepackage{type1cm}        
\usepackage{makeidx}         
\usepackage{graphicx}        
\usepackage{multicol}        
\usepackage[bottom]{footmisc}

\usepackage{amsmath,amssymb}

\makeindex             

\newcommand{\newatop}[2]{\genfrac{}{}{0pt}{}{#1}{#2}}

\newcommand{\bb}[1]{{\mathbb{#1}}}

\newcommand{\puno}{{\textbf{u}}}
\newcommand{\muno}{\mathbf{d}}
\newcommand{\chess}{\mathbf{c}}
\newcommand{\spazio}{X}

\newlength{\pecettawidth}
\begin{document}

\title*{Basic Ideas to Approach Metastability 
        in Probabilistic Cellular Automata}
\author{Emilio N.M.\ Cirillo, Francesca R.\ Nardi and Cristian Spitoni}
\institute{Emilio N.M. Cirillo \at Dipartimento di Scienze di Base e 
           Applicate per l'Ingegneria, 
           Sapienza Universit\`a di Roma, 
           via A.\ Scarpa 16, I--00161, Roma, Italy,
           \email{emilio.cirillo@uniroma1.it}
\and 
Francesca R.\ Nardi \at
Department of Mathematics and Computer Science,
Eindhoven University of Technology,
P.O.\ Box 513, 5600 MB Eindhoven, The Netherlands \at
Eurandom, P.O.\ Box 513, 5600 MB, Eindhoven, The Netherlands,
\email{F.R.Nardi@tue.nl}
\and
Cristian Spitoni \at
Institute of Mathematics,
University of Utrecht, Budapestlaan 6, 3584 CD Utrecht, The~Netherlands,
\email{C.Spitoni@uu.nl}
}
%
%
\maketitle

\abstract*{Cellular Automata are discrete--time dynamical systems on a
spatially extended discrete space which
provide paradigmatic examples of
nonlinear phenomena.
Their stochastic generalizations, i.e.,
Probabilistic Cellular Automata,
are discrete time Markov chains
on lattice with finite single--cell states whose
distinguishing feature is the \textit{parallel} character of the updating rule.
We review the some of the results obtained about the metastable
behavior of Probabilistic Cellular Automata and we try to
point out difficulties and peculiarities with respect to
standard Statistical Mechanics Lattice models.}

\abstract{Cellular Automata are discrete--time dynamical systems on a
spatially extended discrete space which
provide paradigmatic examples of
nonlinear phenomena.
Their stochastic generalizations, i.e.,
Probabilistic Cellular Automata,
are discrete time Markov chains
on lattice with finite single--cell states whose
distinguishing feature is the \textit{parallel} character of the updating rule.
We review the some of the results obtained about the metastable
behavior of Probabilistic Cellular Automata and we try to
point out difficulties and peculiarities with respect to
standard Statistical Mechanics Lattice models.}

\section{Introduction}
\label{s:introduzione}
Cellular Automata are discrete--time dynamical systems on a 
spatially extended discrete space. 
They are well known for being easy to 
implement and
for exhibiting a rich and complex nonlinear behavior as 
emphasized for instance in~\cite{Wolfram1984Nature} 
for Cellular Automata on one--dimensional lattice. 
For the general theory of deterministic Cellular Automata
we refer to the recent
paper~\cite{Kari} and references therein.

Probabilistic Cellular Automata (PCA) are Cellular 
Automata straightforward generalization 
where the updating rule is stochastic.
They are used as models in 
a wide range of applications.
From a theoretic perspective, the main challenges 
concern the non--ergodicity of these dynamics for an infinite collection
of interacting cells. 

Strong relations exist between PCA and the general equilibrium statistical 
mechanics framework~\cite{Wolfram:RevModPhys.55.601,lms}. 
Important issues are related to the interplay between disordered global states 
and ordered phases (\textit{emergence of organized global states, phase 
transition})~\cite{PCA:order:disorder}. 
Although, PCA initial interest arose in the framework of 
Statistical Physics, in the recent literature many different applications
of PCA have been proposed. In particular it is notable to remark 
that a natural context in which the PCA main ideas are of interest 
is that of evolutionary games \cite{PG}.

In this paper we shall consider a particular class of PCA, called 
\textit{reversible} PCA, which are reversible with respect to 
a Gibbs--like measure defined via a translation invariant 
multi--body potential. 
In this framework we shall pose the problem of metastability and 
show its peculiarities in the PCA world.

Metastable states are ubiquitous in nature and are characterized 
by the following phenomenological properties: 
(i) the system exhibits a single phase different from the equilibrium 
predicted by thermodynamics.
The system obeys the usual laws of thermodynamics if 
small variations of the thermodynamical parameters (pressure, 
temperature, ...) are considered. 
(ii) If the system is isolated the equilibrium state is reached after 
a very large random time; the life--time of the metastable 
state is practically infinite. 
The exit from the metastable state can be made easier by forcing 
the appearance large fluctuations of the stable state (droplets of 
liquid inside the super--cooled vapor, ...).
(iii)
The exit from the metastable phase is irreversible.

The problem of the rigorous mathematical description of metastable 
states has long history which started in the 70s, blew up 
in the 90s, and is still an important topic of mathematical 
literature. Different theories have been proposed and developed and 
the pertaining literature is huge. We refer the interested reader 
to the monograph \cite{OV}. In this paper we shall focus on the 
study of metastability in the framework of PCA.

In~\cite{BCLS,CN,CNS1,CNS2,NS} the  
metastable behavior of a certain class of reversible PCA has been analyzed.
In this framework it has been pointed out the remarkable 
interest of a particular reversible PCA
(see Section~\ref{s:croce}) characterized by the fact that the
updating rule of a cell 
depends on the status of the five cells forming a cross 
centered at the cell itself. 
In this model, the future state of the spin at a given cell 
depends also on the present value of such a spin. This 
effect will be called \textit{self--interaction} and its weight 
in the updating rule 
will be called \textit{self--interaction intensity}.

The paper is organized as follows. 
In Section~\ref{s:rpca-sezione} we introduce 
reversible Probabilistic Cellular Automata and discuss some general 
properties. 
In Section~\ref{s:croce} we introduce the model that will be studied 
in this paper, namely, the nearest neighbor and the cross PCA, and discuss 
its Hamiltonian. 
In Section~\ref{s:meta-ingr} we pose the problem of metastability in the 
framework of Probabilistic Cellular Automata and describe 
the main ingredients that are necessary for a full description of 
this phenomenon. In Section~\ref{s:meta-res} we finally 
state our results. 

\section{Reversible Probabilistic Cellular Automata}
\label{s:rpca-sezione}
We shall first briefly recall the 
definition of 
Probabilistic Cellular Automata and then introduce the so called 
\textit{Reversible Probabilistic Cellular Automata}.

Let $\Lambda\subset\bb{Z}^d$ be a finite cube with periodic boundary 
conditions.
Associate with each site $i\in\Lambda$ (also called \textit{cell})
the \text{state variable}
$\sigma_i\in\spazio_0$, 
where $\spazio_0$ is a finite single--site space and denote by 
$\spazio:=\spazio_0^\Lambda$ the \textit{state space}.
Any $\sigma\in\spazio$ is called a \textit{state} or \textit{configuration}
of the system. 

\begin{figure}[t]
\sidecaption
\setlength{\unitlength}{2.3pt}
 \begin{picture}(10,55)(45,-5)
 \thinlines
 \multiput(-5,0)(0,10){5}{\put(0,0){\line(1,0){50}}}
 \multiput(0,-5)(10,0){5}{\put(0,0){\line(0,1){50}}}
 \put(-12,20){${\Lambda}$}
 \put(10,20){\circle*{2}}
 \put(11,21){${0}$}
 \thicklines
 \put(5,15){\line(0,1){20}}
 \put(5,15){\line(1,0){10}}
 \put(5,35){\line(1,0){20}}
 \put(15,15){\line(0,1){10}}
 \put(15,25){\line(1,0){10}}
 \put(25,25){\line(0,1){10}}
 \put(1,13){${I}$}
 \put(30,0){\circle*{2}}
 \put(31,1){${i}$}
 \put(25,-5){\line(0,1){20}}
 \put(25,-5){\line(1,0){10}}
 \put(25,15){\line(1,0){20}}
 \put(35,-5){\line(0,1){10}}
 \put(35,5){\line(1,0){10}}
 \put(45,5){\line(0,1){10}}
 \put(47,14){${i+I}$}
 \end{picture}
 \caption{Schematic representation of the action of the 
          shift $\Theta_{i}$ defined in \eqref{shift}.}
 \label{f:def}
\end{figure}
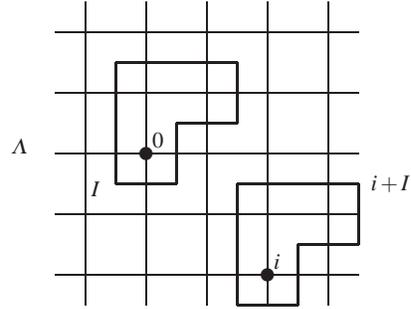

We introduce the shift $\Theta_i$ on the torus, for any $i\in\Lambda$, 
defined as the map 
$\Theta_i:\spazio\to\spazio$ 
shifting a configuration in $\spazio$ so that the site $i$ is mapped to the 
origin $0$, more precisely such that (see figure~\ref{f:def})
\begin{equation}
\label{shift}
(\Theta_i\sigma)_j=\sigma_{i+j}.
\end{equation}
The configuration $\sigma$ at site $j$ shifted by $i$ 
is equal to the configuration at site $i+j$. 
For example (see figure~\ref{f:def}) set $j=0$, then 
the value of the spin at the origin $0$ will be mapped to site $i$. 

We consider a
probability distribution
$f_{\sigma}:\spazio_0\to[0,1]$ depending
on the state $\sigma$ restricted to $I\subset\Lambda$.
A Probabilistic Cellular Automata is 
the Markov chain $\sigma(0),\sigma(1),\dots,\sigma(t)$
on $\spazio$ with transition matrix
\begin{equation}
\label{regola-pca}
p(\sigma,\eta)=\prod_{i\in\Lambda}f_{\Theta_{i}\sigma} (\eta_i)
\end{equation}
for $\sigma,\eta\in\spazio$. We remark that  $f$ depends on $\Theta_{i}\sigma$ only via the neighborhood $i+I$. 
Note that the character of the evolution is local and parallel:
the probability that the spin at the site $i$ assumes at time $t+1$ 
the value $s\in\spazio_0$
depends on the
value of the state variables at time $t$ (parallel evolution)
associated only with the sites in $i+I$ (locality).

A class of \textit{reversible} PCA can be obtained by choosing 
$\spazio=\{-1,+1\}^\Lambda$,
and probability distribution
\begin{equation}
\label{frev}
 f_{\sigma}(s)=\frac{1}{2}\Big\{
        1+s\tanh\Big[\beta\Big(\sum_{j\in\Lambda}k(j)\sigma_j+h\Big)\Big]\Big\}
\end{equation}
for all $s\in\{-1,+1\}$
where $T\equiv1/\beta>0$ 
and $h\in\mathbb{R}$ are called \textit{temperature} and 
\textit{magnetic field}. The function $k:\bb{Z}^2\to\bb{R}$ is such that its 
support\footnote{Recall that, by definition, the support of the 
function $k$ is the subset of $\Lambda$ where the function~$k$ is 
different from zero.}
is a subset of~$\Lambda$ and~$k(j)=k(j')$
whenever $j,j'\in\Lambda$ are symmetric with respect to the origin. 
With the notation introduced above, the set~$I$ is the support 
of the function~$k$.
We shall denote by $p_{\beta,h}$ the corresponding transition matrix 
defined by \eqref{regola-pca}.

Recall that $\Lambda$ is a finite torus, namely, periodic boundary 
conditions are considered throughout this paper. 
It is not difficult to prove~\cite{GJH,KV}
that the above specified PCA dynamics is reversible 
with respect to the finite--volume Gibbs--like 
measure
\begin{equation}
\label{gibbs}
 \mu_{\beta,h}(\sigma)
 =
 \frac{1}{Z_{\beta,h}}
\,e^{-\beta G_{\beta,h}(\sigma)}
\end{equation}
with \textit{Hamiltonian}
\begin{equation}
\label{mahG}
G_{\beta,h}(\sigma)
=
 -h\sum_{i\in\Lambda}\sigma_i
 -\frac{1}{\beta}\sum_{i\in\Lambda}
    \log\cosh\Big[\beta
   \Big(\sum_{j\in\Lambda}k(j-i)\sigma_j+h\Big)\Big]
\end{equation}
and \textit{partition function}
$ Z_{\beta,h}= \sum_{\eta\in\spazio} \exp\{-\beta G_{\beta,h}(\eta)\}$.
In other words, in this case the detailed balance equation
\begin{equation}
\label{dett}
p_{\beta,h}(\sigma,\eta) e^{-\beta G_{\beta,h}(\sigma)}
=e^{-\beta G_{\beta,h}(\eta)} p_{\beta,h}(\eta,\sigma)
\end{equation}
is satisfied thus the probability measure~$\mu_{\beta,h}$ is stationary for 
the PCA. 

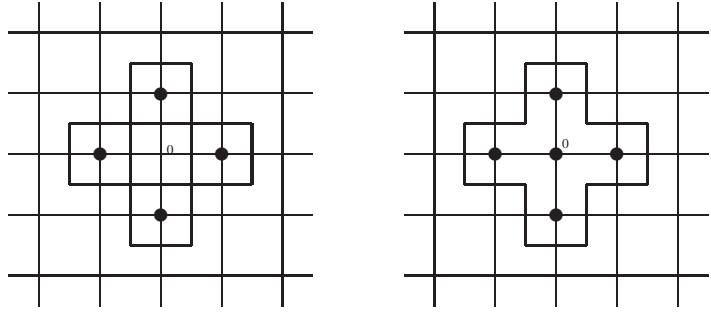
\begin{figure}[t]
\setlength{\unitlength}{2.3pt}
 \begin{picture}(10,55)(-20,-10)
 \thinlines
 \multiput(-5,0)(0,10){5}{\put(0,0){\line(1,0){50}}}
 \multiput(0,-5)(10,0){5}{\put(0,0){\line(0,1){50}}}
 \put(20,30){\circle*{2}}
 \put(30,20){\circle*{2}}
 \put(20,10){\circle*{2}}
 \put(10,20){\circle*{2}}
 \put(21,20){${\scriptscriptstyle 0}$}
 \thicklines
 \put(5,15){\line(0,1){10}}
 \put(15,15){\line(0,1){10}}
 \put(25,15){\line(0,1){10}}
 \put(15,15){\line(1,0){10}}
 \put(5,25){\line(1,0){10}}
 \put(15,25){\line(0,1){10}}
 \put(15,25){\line(1,0){10}}
 \put(15,35){\line(1,0){10}}
 \put(25,35){\line(0,-1){10}}
 \put(25,25){\line(1,0){10}}
 \put(35,25){\line(0,-1){10}}
 \put(35,15){\line(-1,0){10}}
 \put(25,15){\line(0,-1){10}}
 \put(25,5){\line(-1,0){10}}
 \put(15,5){\line(0,1){10}}
 \put(15,15){\line(-1,0){10}}
 \thinlines
 \multiput(60,0)(0,10){5}{\put(0,0){\line(1,0){50}}}
 \multiput(65,-5)(10,0){5}{\put(0,0){\line(0,1){50}}}
 \put(85,30){\circle*{2}}
 \put(95,20){\circle*{2}}
 \put(85,10){\circle*{2}}
 \put(75,20){\circle*{2}}
 \put(85,20){\circle*{2}}
 \put(86,21){${\scriptscriptstyle 0}$}
 \thicklines
 \put(70,15){\line(0,1){10}}
 \put(70,25){\line(1,0){10}}
 \put(80,25){\line(0,1){10}}
 \put(80,35){\line(1,0){10}}
 \put(90,35){\line(0,-1){10}}
 \put(90,25){\line(1,0){10}}
 \put(100,25){\line(0,-1){10}}
 \put(100,15){\line(-1,0){10}}
 \put(90,15){\line(0,-1){10}}
 \put(90,5){\line(-1,0){10}}
 \put(80,5){\line(0,1){10}}
 \put(80,15){\line(-1,0){10}}
 \end{picture}
\caption{Schematic representation of the nearest neighbor (left) 
         and cross (right) models.}
\label{f:modelli}
\end{figure}

Note that different reversible PCA models can be specified by 
choosing different functions~$k$. In particular the support~$I$ of such 
a function can be varied. 
In the next section we shall introduce two 
common choices, the \textit{nearest neighbor PCA}~\cite{CN}
obtained by choosing the support of~$k$ as the set of the four sites 
neighboring the 
origin and the \textit{cross PCA}~\cite{CNS2}
obtained by choosing the support of~$k$ as the set made of the origin and its 
four neighboring sites (see figure~\ref{f:modelli}).

The stationary measure $\mu_{\beta,h}$ introduced above looks like a 
finite--volume Gibbs measure with Hamiltonian
$G_{\beta,h}(\sigma)$ (see \eqref{mahG}). It is worth noting that 
$G_{\beta,h}$ cannot be thought as 
a proper statistical mechanics Hamiltonian since it depends on the temperature 
$1/\beta$. On the other hand the 
low--temperature behavior of the stationary measure of the PCA 
can be guessed by looking at the \textit{energy} function
\begin{equation}
\label{zth}
H_h(\sigma)
=
\lim_{\beta\to\infty}G_{\beta,h}(\sigma)
=
-h\sum_{i\in\Lambda}\sigma_i
-\sum_{i\in\Lambda}
   \Big|\sum_{j\in\Lambda}k(j-i)\sigma_j+h\Big|
\end{equation}
The absolute minima of the function $H_h$ are called 
\textit{ground states} of the stationary measure for the reversible PCA.

\section{The tuned cross PCA}
\label{s:croce}
We consider, now, a particular example of reversible PCA. 
More precisely, we set 
$k(j)=0$ if $j$ is neither the origin
nor one of its nearest neighbors, i.e., it is not in the five site cross 
centered at the origin, $k(0)=\kappa\in[0,1]$, and $k(j)=1$ if $j$ is one of 
the four nearest neighbor of the origin; we shall denote by $J$ 
the set of nearest neighbors of the origin.
With such a choice we have that
\begin{equation}
\label{tcross}
 f_{\sigma}(s)=\frac{1}{2}\Big\{
   1+s\tanh\Big[\beta\Big(
   \kappa\sigma_0+\sum_{j\in J}\sigma_j+h
                          \Big)\Big]\Big\}
=
\frac{1}
     {1+e^{-2\beta s (\kappa\sigma_0+\sum_{j\in J}\sigma_j+h)}}
\end{equation}
We shall call this model the \textit{tuned cross} PCA.
The \textit{self--interaction intensity}
$\kappa$ tunes 
between 
the \textit{nearest neighbor} $(\kappa=0)$ and 
the \textit{cross} $(\kappa=1)$ PCA. 

Note that for this model the Hamiltonian $G_{\beta,h}$ 
defining the stationary 
Gibbs--like measure is given by 
\begin{equation}
\label{ham-cross}
G_{\beta,h}(\sigma)
=
 -h\sum_{i\in\Lambda}\sigma_i
 -\frac{1}{\beta,h}\sum_{i\in\Lambda}
    \log\cosh\Big[\beta
   \Big(
   \kappa\sigma_i+\sum_{j\in i+J}\sigma_j +h \Big)\Big]
\end{equation}
while the corresponding energy function, see \eqref{zth}, is 
\begin{equation}
\label{zth-cross}
H_h(\sigma)
=
-h\sum_{i\in\Lambda}\sigma_i
-\sum_{i\in\Lambda}
   \Big|\kappa\sigma_i+\sum_{j\in i+J}\sigma_j+h\Big|
\end{equation}

\begin{figure}[t]
\sidecaption
\begin{picture}(10,90)(60,10)
\thinlines
\multiput(-20,0)(20,0){5}{\line(0,1){90}}
\multiput(-25,5)(0,20){5}{\line(1,0){90}}
\thicklines
\put(-20,25){\circle*{5}}
\put(0,45){\circle*{5}}
\put(-20,25){\line(1,1){20}}
\put(0,65){\circle*{5}}
\put(40,65){\circle*{5}}
\qbezier(0,65)(20,75)(40,65)
\put(20,25){\circle*{5}}
\put(40,45){\circle*{5}}
\put(40,5){\circle*{5}}
\put(60,25){\circle*{5}}
\put(20,25){\line(1,1){20}}
\put(20,25){\line(1,-1){20}}
\put(60,25){\line(-1,1){20}}
\put(60,25){\line(-1,-1){20}}
\put(-20,85){\circle*{5}}
\put(-20,5){\circle*{5}}
\put(0,25){\circle*{5}}
\put(20,5){\circle*{5}}
\put(-20,5){\line(1,1){20}}
\put(-20,5){\line(1,0){40}}
\put(0,25){\line(1,-1){20}}
\end{picture}
\caption{Schematic representation of the coupling constants: from the left to
the right and from the top to the bottom the couplings 
$J_{.}$, 
$J_{\langle\langle\langle \rangle\rangle\rangle}$, 
$J_{\langle\langle\rangle\rangle}$,
$J_{\triangle}$, and
$J_{\diamondsuit}$
are depicted.}
\label{f:accoppiamenti}
\end{figure}
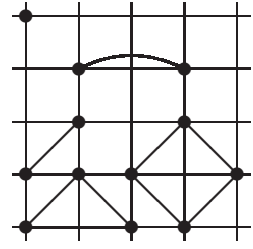

In statistical mechanics lattice systems, the energy of a configuration 
is usually written in terms of coupling constants.
We could write the expansion of the energy $H_h$ in \eqref{zth-cross}, 
but, for the sake of simplicity, we consider the 
nearest neighbor PCA \cite{CN}, namely, we set $\kappa=0$. We get
\begin{displaymath}
\begin{array}{ll} 
{\displaystyle{ 
H_h(\sigma)=}} 
& 
{\displaystyle{ 
 \!\!\!
-J_{.} 
\sum_{x\in\Lambda}\sigma(x) 
-J_{_{\langle\langle\rangle\rangle}} 
\sum_{\langle\langle x y\rangle\rangle}\sigma(x)\sigma(y) 
-J_{_{\langle\langle\langle \rangle\rangle\rangle}} 
\sum_{\langle\langle\langle x y\rangle\rangle\rangle}\sigma(x)\sigma(y) 
}}\\ 
& 
{\displaystyle{ 
 \!\!\!
-J_{_{\triangle}} 
\sum_{\triangle_{xyz}}\sigma(x)\sigma(y)\sigma(z) 
-J_{_{\diamondsuit}} 
\sum_{\diamondsuit_{xywz}}\sigma(x)\sigma(y)\sigma(w)\sigma(z)  
}} 
\\ 
\end{array} 
\end{displaymath}
where the meaning of the symbols $\cdot$, 
$\langle\langle\rangle\rangle$,
$\langle\langle\langle\rangle\rangle\rangle$
$\triangle$, and $\diamondsuit$ 
is illustrated in figure~\ref{f:accoppiamenti} and the 
corresponding coupling constants are 
\begin{displaymath} 
  J_{.} = \frac{5}{2} h, \,
  J_{_{\langle\langle\rangle\rangle}}= 1-\frac{1}{4}h,\,
  J_{_{\langle\langle\langle \rangle\rangle\rangle}}
      =\frac{1}{2}-\frac{1}{8}h,\,
  J_{_{\triangle}} = -\frac{1}{8}h,\,
  \textrm{ and }
 J_{_{\diamondsuit}}
    = -\frac{1}{2}+\frac{3}{8} h
\end{displaymath} 
It is interesting to note that the coupling constant $J_\diamondsuit$ 
is negative (antiferromagnetic coupling), this will give a physical meaning 
to the appearance of checkerboard configurations in the study of 
metastability for the nearest neighbor PCA. 

The coupling constants can be computed by using 
\cite[equations~(6) and (7)]{CS96} 
(see also \cite[equations~(3.1) and (3.2)]{HK} and \cite{CNP}). 
More precisely, given $f:\{-1,+1\}^V\to\bb{R}$, with $V\subset\bb{Z}^2$ 
finite, we have that for any $\sigma\in\{-1,+1\}^V$
\begin{equation}
\label{pot000}
f(\sigma)
=
\sum_{I\subset V} C_I\prod_{i\in I}\sigma_i
\end{equation}
with the coefficients $C_I$'s given by 
\begin{equation}
\label{pot010}
C_I
=
\frac{1}{2^{|V|}}
\sum_{\sigma\in\{-1,+1\}^V}f(\sigma)\prod_{i\in I}\sigma_i
\end{equation}
We refer to \cite{CLRS} for the details. We note that in that paper 
the couplings have been computed for a more general model than the one 
discussed here. 

Now, we jump back to the tuned cross PCA and we discuss the structure of 
the ground states, that is to say, we study the global minima of the 
energy function $H_h$ given in \eqref{zth-cross}.
Such a function can be rewritten as 
\begin{displaymath}
H_h(\sigma)
=
 \sum_{i\in\Lambda}
 H_{h,i}(\sigma)
\end{displaymath}
with
\begin{equation}
\label{zth-cross-010}
H_{h,i}(\sigma)
=
-\Big[\frac{1}{5}h\Big(\sigma_i+\sum_{j\in i+J}\sigma_j\Big)
   +\Big|\kappa\sigma_i+\sum_{j\in i+J}\sigma_j+h\Big|
  \Big]
\end{equation}
We also note that 
\begin{equation}
\label{zth-cross-020}
H_h(\sigma)=H_{-h}(-\sigma)
\end{equation}
for any $h\in\bb{R}$ and $\sigma\in\spazio$, where $-\sigma$ denotes 
the configuration obtained by flipping the sign of all the spins of $\sigma$.
By \eqref{zth-cross-020} we can bound our discussion to the case $h\ge0$ and
deduce a posteriori the structure of the ground states for $h<0$. 

\begin{figure}[t]
\sidecaption
\setlength{\unitlength}{0.8pt}
\begin{picture}(200,120)(20,-10)  
\qbezier(100,0)(100,50)(100,100)
\put(200,50){\vector(1,0){5}}
\put(100,100){\vector(0,1){5}}
\thicklines
\put(100,50){\line(1,0){100}}
\put(100,50){\circle*{7}}
\put(200,35){${k_0}$}
\put(105,100){${h}$}
\put(140,70){${\puno}$}
\put(140,20){${\muno}$}
\thinlines
\qbezier(20,50)(75,75)(92,58)
\put(90,60){\vector(1,-1){5}}
\put(-20,30){${\puno}$,}
\put(0,30){${\muno}$,}
\put(20,30){${\chess_\textrm{e}}$,}
\put(40,30){${\chess_\textrm{o}}$}
\end{picture}
\caption{Zero temperature phase diagram of the stationary measure of the 
tuned cross PCA.
On the thick lines the ground states of the adjacent
regions coexist. At the origin the listed four ground states coexist.}
\label{f:ground}
\end{figure}
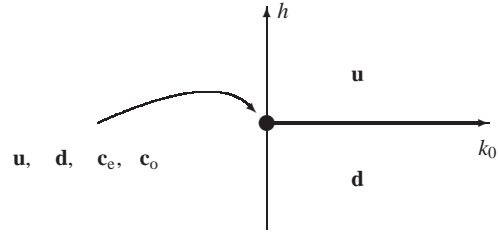

The natural candidates to be ground states are the following configurations: 
$\puno\in\spazio$ such that $\puno(i)=+1$ for all $i\in\Lambda$, 
$\muno\in\spazio$ such that $\muno(i)=+1$ for all $i\in\Lambda$, 
$\chess_\textrm{e}$, and $\chess_\textrm{o}$ with 
$\chess_\textrm{e}$ the checkerboard configuration with pluses on the 
even sub--lattice of $\Lambda$ and minuses on its complement, while 
$\chess_\textrm{o}$ is the corresponding spin--flipped configuration.
Indeed, we can prove that the structure of the zero--temperature phase 
diagram is that depicted in figure~\ref{f:ground}.

\par\noindent\textit{Case $h>0$ and $k_0\ge0$.\/}
The minimum of $H_{h,i}$ is attained at the cross configuration having 
all the spins equal to plus one. Hence the unique absolute minimum 
of $H_h$ is the state
$\puno$.

\par\noindent\textit{Case $h=0$ and $k_0>0$.\/}
The minimum of 
\begin{displaymath}
H_{0,i}(\sigma)
=-\Big|\kappa\sigma_i+\sum_{j\in i+J}\sigma_j\Big|
\end{displaymath}
is attained at the cross configuration having 
all the spins equal to plus one or all equal to minus one. 
Hence the set of ground states is made of the two configurations 
$\puno$ and $\muno$.

\par\noindent\textit{Case $h=0$ and $k_0=0$.\/}
The minimum of $H_{0,i}$ is attained at the cross configuration having 
all the spins equal to plus one or all equal to minus one on the neighbors 
of the center and with the spin at the center which can be, in any case, 
either plus or minus. 
Hence the set of ground states is made of the four configurations 
$\puno$, $\muno$, $\chess_\textrm{e}$, and $\chess_\textrm{o}$.

\par\noindent\textit{Case $h<0$.\/} The set of ground states 
can be easily discussed as for $h>0$ by using the 
property \eqref{zth-cross-020}.

\section{Main ingredients for metastability}
\label{s:meta-ingr}
At $\kappa>0$, 
the zero temperature phase diagram in figure~\ref{f:ground}
is very similar to that of the standard Ising model, which is 
the prototype for the description of phase transitions in 
Statistical Mechanics. So we expect that even in the case of the 
tuned cross PCA 
the equilibrium behavior could be described as follows: (i) at positive 
magnetic field $h$ there exist a unique phase with positive 
magnetization\footnote{By exploiting the translational invariance of the model,
it is possible to define the magnetization as the mean value 
of the spin at the origin against the Gibbs--like 
equilibrium measure $\mu_{\beta,h}$.}; (ii) the same it is true at negative 
$h$ but with negative magnetization; (iii) at $h=0$ the 
equilibrium behavior is more complicated: there exists a critical value of 
the temperature such that at temperatures larger than such a value 
there exists a unique phase with zero magnetization, while at 
temperatures smaller than the critical one there exists two 
equilibrium measures with opposite not zero magnetization, called the 
\textit{residual magnetization}.

This scenario has proven to be true in the case of the two--dimensional
standard Ising model, but in the context of the tuned cross PCA 
the problem is much more difficult due to the complicated structure of 
the energy function \eqref{ham-cross}. The validity of such a scenario 
has been checked via a Mean Field computation in \cite{CLRS}.

From now on, for technical reasons, we shall assume that the magnetic field
satisfies the following conditions
\begin{equation}
\label{cm}
0<h<4
\;\;\;\textrm{ and }\;\;\;
h\neq\kappa,2-\kappa,2+\kappa,4-\kappa,4+\kappa
\end{equation}
Since $h>0$, the equilibrium is characterized by 
positive magnetization. The question is: is it possible to investigate 
the possibility of the existence of metastable states? 
In other words, is it possible to show that there exist not 
equilibrium phases in which the system is trapped in the sense described 
in the introduction (see Section~\ref{s:introduzione})?

This question has a very long history: in some sense it arose
with the van der Waals theory of liquid--vapor transition and 
began to find some mathematically rigorous answer only in the 
80's. We just quote \cite{OS} for the \textit{pathwise approach} 
and \cite{BEGK} for the \textit{potential theoretic} one and 
we refer to \cite{OV} for the full
story and for complete references. 

According to the rigorous theories of metastability 
the problem has to be approached from a dynamical point of view. 
Namely, we shall consider the evolution of the tuned cross PCA started at 
the initial configuration $\zeta\in\spazio$ and study the random 
variable
\begin{equation}
\label{met00}
\tau^\zeta_\puno:=\inf\{t>0,\,\sigma(t)=\puno\}
\end{equation}
called the \textit{first hitting time} to $\puno$. 
The state $\zeta$ will be called metastable or not depending on the properties 
of the random variable $\tau^\zeta_\puno$ in the zero temperature 
limit\footnote{The regime outlined in this paper, i.e., finite 
state space and temperature tending to zero, is usually called the 
Wentzel--Friedlin regime. Different limits can be considered, 
for instance, volume tending to infinity.}
($\beta\to\infty$). In the framework of different 
approaches to metastability 
different definitions of metastable states have been given, but 
they are all related to the properties of the hitting time $\tau^\zeta_\puno$.
In particular it has to happen that the mean value of $\tau^\zeta_\puno$
has to be large, say diverging exponentially fast with $\beta\to\infty$.

As remarked above, for $h>0$ small, natural candidates to be metastable 
states for the tuned cross PCA are the configurations 
$\muno$, $\chess_\textrm{e}$, and $\chess_\textrm{o}$. 
But, imagine to start the chain at $\muno$: why should such a state 
be metastable? Why should the chain take a very long time to hit 
the ``stable" state $\puno$? The analogous question posed in the 
framework of the two--dimensional Ising model with 
Metropolis dynamics has an immediate
qualitative answer: in order to reach $\puno$ starting from $\muno$
the system has to perform, spin by spin, a sequence of 
changes against the energy drift. Indeed, plus spins have to 
be created in the starting sea of minuses, and those transitions 
have a positive energy cost if the magnetic field is small enough, 
indeed the interaction is ferromagnetic and pairs of neighboring 
opposite spins have to be created. 

But in the case of the tuned cross PCA, recall \eqref{zth-cross} 
and recall we assumed $h<4$, see \eqref{cm},
the starting $\muno$ and the final $\puno$ configurations have energy 
\begin{displaymath}
H_h(\muno)=-|\Lambda|(4+\kappa-2h)
\;\;\textrm{ and }\;\;
H_h(\puno)=-|\Lambda|(4+\kappa+2h)
\end{displaymath}
So that $H_h(\muno)>H_h(\puno)$, as it is obvious since 
$\puno$ is the ground state. Moreover, the dynamics is allowed 
to jump in a single step from $\muno$ to $\puno$ by reversing 
all the spins of the system. 
A naive (wrong) conclusion would be that $\muno$ cannot be metastable 
because the jump from $\muno$ to $\puno$ can be performed 
in a single step by decreasing the energy. 

The conclusion is wrong because in reversible PCA the probability 
to perform a jump is not controlled simply by the difference of energies
of the two configurations involved in the jump. Indeed, in the 
example discussed above, recall \eqref{regola-pca} and \eqref{tcross}, 
we have that 
\begin{displaymath}
p_{\beta,h}(\muno,\puno)=
\Big[\frac{1}{1+e^{2\beta(4+\kappa-h)}}\Big]^{|\Lambda|}
\stackrel{\beta\to\infty}{\sim}
e^{-2|\Lambda|\beta(4+\kappa-h)}
\end{displaymath}
which proves that the direct jump from $\muno$ to $\puno$ is 
depressed in probability when $\beta$ is large. 

This very simple remark shows that the behavior of the PCA 
cannot be analyzed by simply considering the energy difference 
between configurations. It is quite evident that a suitable 
cost function has to be introduced. 

From \eqref{cm} the local field 
$\kappa\sigma_0+\sum_{j\in J}\sigma_j+h$ 
appearing in \eqref{tcross} 
is different from zero.
Thus, for $\beta\to\infty$, 
 \begin{displaymath}
 p_{\beta,h}(\sigma,\eta)
 \to
 \left\{
 \begin{array}{l}
 {\displaystyle
  1\;\;\;\;
   \textrm{ if } \eta(i)\big[\kappa\sigma_i+\sum_{j\in i+J}\sigma_j+h\big]>0 
    \;\;\forall i\in\Lambda
 }\\
 0\;\;\;\;
  \textrm{ otherwise }
 \end{array}
 \right.
 \end{displaymath}
where we have used \eqref{regola-pca}.
Hence, 
given $\sigma$, there exists a unique configuration $\eta$ such that 
$p_{\beta,h}(\sigma,\eta)\to1$ for $\beta\to\infty$ and this 
configuration is the one such that $\eta(i)$ 
is aligned with the local field 
$\kappa\sigma_i+\sum_{j\in i+J}\sigma_j+h$ for any $i\in\Lambda$.
Such a unique configuration will be called 
the \textit{downhill image} of $\sigma$.
This property explains well in which sense PCA are the probabilistic 
generalization of deterministic Cellular Automata: indeed, in such models 
each configuration is changed deterministically into a unique 
image configuration. This property is recovered in probability 
in reversible PCA in the limit $\beta\to\infty$.

We now remark that if $\eta$ is different from 
the downhill image of $\sigma$, we have that 
$p_{\beta,h}(\sigma,\eta)$ decays exponentially with 
rate
\begin{equation}
\label{meta20}
\Delta_h(\sigma,\eta)=
 -\lim_{\beta\to\infty}\frac{1}{\beta}\log p_{\beta,h}(\sigma,\eta)
 =
\!\!\!\!\!\!\!\!\!
\sum_{\newatop{i\in\Lambda:}
     {\eta(i)[\kappa\sigma_i+\sum_{j\in i+J}\sigma_j+h]<0}}
\!\!\!
2\Big|\kappa\sigma_i+\sum_{j\in i+J}\sigma_j+h\Big|
\end{equation}
Note that if $\eta$ is the downhill image of $\sigma$ then 
$\Delta_h(\sigma,\eta)=0$.
More precisely we have 
\begin{displaymath}
e^{-\beta\Delta_h(\sigma,\eta)-\beta\gamma(\beta)}
\le
p_{\beta,h}(\sigma,\eta)
\le
e^{-\beta\Delta_h(\sigma,\eta)+\beta\gamma(\beta)}
\end{displaymath}
with $\gamma(\beta)\to0$ for $\beta\to\infty$.
This property is known in the literature as the Wentzel and Friedlin 
condition. 

Since from \eqref{dett} and \eqref{meta20} it follows that the 
following reversibility condition 
\begin{equation}
\label{rev}
H_h(\sigma)+\Delta_h(\sigma,\eta)
=
H_h(\eta)+\Delta_h(\eta,\sigma)
\end{equation}
is satisfied for any $\sigma,\eta\in\spazio$, we have that the 
function  
$\Delta_h(\sigma,\eta)$ can be interpreted as the 
the energy cost 
that must be paid in the transition $\sigma\to\eta$.

We are now ready to give a precise definition of metastable 
states in the framework of reversible Probabilistic Cellular Automata.
We shall follow the approach in \cite{MNOS} which is based on 
the analysis of the energy landscape of the system. Note that 
in our setup the energy landscape is not only given by the energy 
function $H_h$, but it is also decorated by the energy cost 
function $\Delta_h$. 
it is important to remark that, for the sake of clearness, 
we shall give the definition having 
in mind the specific case we are considering, namely, the tuned cross 
PCA with $0<h<\kappa$, but the definition we shall can give 
can be easily generalized to the broad context of reversible PCA.  

\begin{figure}[t]
\sidecaption
 \setlength{\unitlength}{0.03cm}
 \begin{picture}(50,150)(240,0)
 \qbezier(130,10)(140,45)(150,80)
 \put(130,10){\circle*{3}}
 \put(132,3){{$\omega_1$}}
 \qbezier(150,80)(180,185)(190,100)
 \put(150,80){\circle*{3}}
 \put(152,73){{$\omega_2$}}
 \qbezier(190,100)(210,160)(220,90)
 \put(190,100){\circle*{3}}
 \put(191,93){{$\omega_3$}}
 \qbezier(220,90)(240,150)(250,80)
 \put(220,90){\circle*{3}}
 \put(221,83){{$\omega_4$}}
 \put(250,80){\circle*{3}}
 \put(251,73){{$\omega_5$}}
 \put(177,50){\vector(0,1){84}}
 \put(165,36){{$\Phi_\omega-H_h(\omega_1)$}}
 \put(177,30){\vector(0,-1){20}}
 \end{picture}
\caption{Graphic representation of the definition of height of a path.}
\label{f:hpath}
\end{figure}
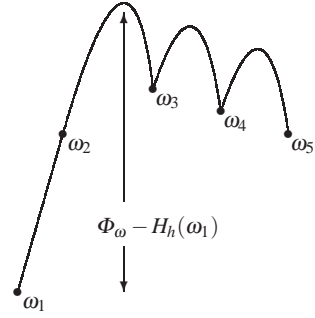

A sequence of configurations $\omega=\{\omega_1,\dots,\omega_n\}$, 
with $\omega_i\in\spazio$ for $i=1,\dots,n$, is called 
\textit{path}. The \textit{height} of the path $\omega$
is defined as 
\begin{equation}
\label{hpath}
\Phi_\omega=\max_{i=1,\dots,n-1}[H_h(\omega_i)+\Delta_h(\omega_i,\omega_{i+1})]
\end{equation}
see figure~\ref{f:hpath} for a graphic illustration.

Given two sets of configurations
$A,A'\subset\spazio$, the  
\textit{communication height} $\Phi(A,A')$ between $A,A'$ 
is defined as 
 \begin{equation}
\label{hcomm}
 \Phi(A,A')=\min_{\omega:A\to A'}\Phi_\omega
 \end{equation}
where the minimum is taken on the set of paths starting in $A$ and 
ending in $A'$.
Given $\sigma\in\spazio$, we define the \textit{stability level} 
of $\sigma$ as
 \begin{equation}
 \label{stability}
 V_\sigma=\Phi(\sigma,\{\textrm{states 
                        with energy smaller than }\sigma\})
          -H_h(\sigma)
 \end{equation}
That is to say, $V_\sigma$ is the height of the most convenient path 
that one has to follow in order to decrease the energy starting from 
$\sigma$. 

Finally, we define the \textit{maximal stability level} as 
the largest among the stability levels, i.e., 
 \begin{equation}
 \label{msl}
 \Gamma_\textrm{m}
 =\max_{\sigma\in\spazio\setminus\{\puno\}}V_\sigma>0
 \end{equation}
and the set of \textit{metastable} states 
\begin{equation}
\label{meta}
\spazio_{\textrm{m}}=
\{\eta\in\spazio\setminus\{\puno\}:\,V_\eta=\Gamma_\textrm{m}\}
\end{equation}

This definition of metastable states is particularly nice, since it is 
based only on the properties of the energy landscape. In other words, 
in order to find the metastable states of the tuned cross PCA, one ``just" 
has to solve some variational problems on the energy landscape  
of the model. This is, unfortunately, a very difficult task that 
has been addressed mainly in \cite{CN,CNS1}. 

Why is this definition of metastable states satisfying? 
Because, given $\zeta\in\spazio_\textrm{m}$, 
for the chain started at $\zeta$, we can prove properties of the 
random variable $\tau^\zeta_\puno$ characterizing $\zeta$ as a metastable 
state in the physical sense outlined in the introduction. 
Indeed, if we let $\bb{P}_\sigma$ and $\bb{E}_\sigma$, respectively, 
the probability and the average computed along the trajectories 
of the tuned cross PCA started at $\sigma\in\spazio$, we can state 
the following theorem.

\medskip
\par\noindent\textbf{Theorem\/}
\textit{Let $\zeta\in\spazio_{\normalfont\textrm{m}}$. For any 
 $\varepsilon>0$ we have that 
 \begin{displaymath}
   \lim_{\beta\to\infty}
     \mathbb{P}_\zeta
       (
        e^{\beta(\Gamma_{\normalfont\textrm{m}}-\varepsilon)}
        <\tau^\zeta_\puno<
        e^{\beta(\Gamma_{\normalfont\textrm{m}}+\varepsilon)}
       )
   =1
 \end{displaymath}
Moreover, 
\begin{displaymath}
\lim_{\beta\to\infty}
\frac{1}{\beta}
 \log\bb{E}_\zeta[\tau^\zeta_\puno]
=\Gamma_{\normalfont\textrm{m}}
\end{displaymath}
}

This theorem has been proven in \cite{MNOS} in the framework of 
Statistical Mechanics lattice systems with Metropolis dynamics. 
Its generalization to the PCA case has been discussed in \cite{CNS1}.

The physical content of the two statements in the theorem is that 
the first hitting time of the chain started at a metastable state 
$\zeta\in\spazio_\textrm{m}$ is of order $\exp\{\beta\Gamma_\textrm{m}\}$.
The first of the two statements ensures this convergence in probability 
and the second in mean. 

It is important to remark that it is possible to give a more detailed 
description of the behavior of the chain started at a metastable state. 
In particular it can be typically proven a nucleation property, that
is to say, one can prove that before touching the stable state $\puno$ 
the chain has to visit ``necessarily" an intermediate configuration 
corresponding to a ``critical" droplet of the stable phase (plus 
one) plunged in the sea of the metastable one. 
By necessarily, above, we mean with probability one in the limit 
$\beta\to\infty$. 
For a wide description of the results that can be proven we refer 
the interested reader, for instance, to \cite{MNOS,OV}.

\section{Metastable behavior of the tuned cross PCA}
\label{s:meta-res}
The metastable behavior of the tuned cross PCA has been 
studied extensively in 
\cite{CN} (nearest neighbor PCA, i.e., $\kappa=0$), 
\cite{BCLS,CNS1} (cross PCA, i.e., $\kappa=1$), and
\cite{CNS2} (tuned cross PCA with $0<\kappa<1$). 
In the extreme cases, i.e., $\kappa=0$ and $\kappa=1$, rigorous results 
were proved, while in the case 
$0<\kappa<1$ only heuristic arguments have been provided.
In this section we shall review briefly the main results 
referring the reader to the quoted papers for details.
We shall always assume that $h$ satisfies \eqref{cm} and 
$2/h$ not integer; moreover, we note that the result listed 
below are proven for $\Lambda$ large enough depending on $h$. 

\begin{figure}[t]
 \setlength{\unitlength}{.07cm}
 \begin{picture}(20,40)(-100,-10)
 \thinlines
 \multiput(-70,0)(0,5){5}{
    \multiput(0,0)(5,0){4}{\put(0,0){${+}$}}
    }
 \put(-50,10){${+}$}
 \qbezier(-40,10)(-25,40)(-10,10)
 \put(-25,16){\vector(0,1){8}}
 \put(-25,8){\vector(0,-1){8}}
 \put(-27,10){${\Gamma_\textrm{m}}$}
 \put(-26,-5){${H(\textbf{d})}$}
 \put(-60,-5){${\textbf{p}_{\ell_{\textrm{c},1}}}$}
 \put(10,-5){${\textbf{p}_{\ell_{\textrm{c},2}}}$}
 \multiput(0,0)(0,5){5}{
    \multiput(0,0)(5,0){4}{\put(0,0){${+}$}}
    }
 \put(20,10){${+}$}
 \put(20,15){${+}$}
 \put(30,16){\vector(0,1){6}}
 \put(30,6){\vector(0,-1){6}}
 \put(28,9){${\ell_{\textrm{c}}=\lfloor\frac{2}{h}\rfloor+1}$}
 \end{picture}
\caption{Graphical description of $\Gamma_\textrm{m}$ for the 
cross PCA.}
\label{f:drop-cr}
\end{figure}
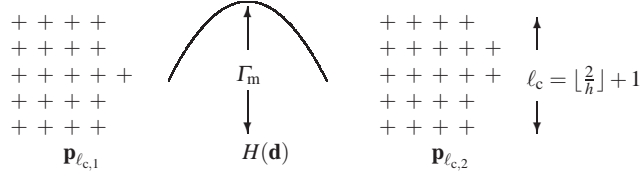

In the cross case ($\kappa=1$) it has been proven \cite{CNS1}
that the metastable state is unique, more precisely, with the 
notation introduced above, it has been shown that 
$\spazio_\textrm{m}=\{\muno\}$. 
Moreover, it has also been proven that the maximal stability level 
is given by 
\begin{equation}
\label{msl-cr}
\Gamma_\textrm{m}
=H_h(\textbf{p}_{\ell_{\textrm{c},1}})
   +\Delta_h(\textbf{p}_{\ell_{\textrm{c},1}},
             \textbf{p}_{\ell_{\textrm{c},2}})-H_h(\muno)
\stackrel{\beta\to\infty}{\sim}
\frac{16}{h}
\end{equation}
where\footnote{Given a real $r$ we denote by 
$\lfloor r\rfloor$ its integer part, namely, the largest integer 
smaller than $r$.} 
$\ell_\textrm{c}=\lfloor 2/h\rfloor+1$ is called 
\textit{critical length}, 
$\textbf{p}_{\ell_{\textrm{c},1}}$ is a configuration characterized 
by a $\ell_\textrm{c}\times(\ell_\textrm{c}-1)$ 
rectangular droplet of plus spins in the sea of minuses
with a single site protuberance attached to one of the two longest 
sides of the rectangle, and 
$\textbf{p}_{\ell_{\textrm{c},2}}$ is a configuration characterized 
by a $\ell_\textrm{c}\times(\ell_\textrm{c}-1)$ 
rectangular droplet of plus spins in the sea of minuses
with a two site protuberance attached to one of the two longest 
sides of the rectangle (see figure~\ref{f:drop-cr}).

Once the model dependent problems have been solved and the metastable 
state found, the properties of such a state are provided by the general 
Theorem stated in Section~\ref{s:meta-ingr}.
We just want to comment that the peculiar expression of the 
maximal stability level that, we recall, gives the exponential 
asymptotic of the mean exit time, has a deep physical meaning. 
Indeed, it is also proven that during the escape from the 
metastable state $\muno$ to the stable one $\puno$ 
the chain visits with probability tending to one in the limit 
$\beta\to\infty$ the configuration 
$\textbf{p}_{\ell_{\textrm{c},1}}$ and, starting from such a 
configuration, it performs the jump to 
$\textbf{p}_{\ell_{\textrm{c},2}}$.
From the physical point of view this property means that the escape 
from the metastable state is achieved via the nucleation of the 
critical droplet $\textbf{p}_{\ell_{\textrm{c},2}}$.

\begin{figure}[t]
 \setlength{\unitlength}{.07cm}
 \begin{picture}(20,40)(-100,-10)
 \thinlines
 \multiput(-70,0)(0,5){5}{
    \multiput(0,0)(5,0){4}{\put(0,0){${-}$}}
    }
 \multiput(-70,0)(0,10){3}{
    \multiput(0,0)(10,0){2}{\put(0,0){${+}$}}
    }
 \multiput(-65,5)(0,10){2}{
    \multiput(0,0)(10,0){2}{\put(0,0){${+}$}}
    }
 \qbezier(-40,10)(-25,40)(-10,10)
 \put(-25,16){\vector(0,1){8}}
 \put(-25,8){\vector(0,-1){8}}
 \put(-27,10){${\Gamma_\textrm{m}}$}
 \put(-26,-5){${H(\textbf{d})}$}
 \put(-60,-5){${\textbf{c}_{\ell_{\textrm{c}}}}$}
 \put(10,-5){${\textbf{c}_{\ell_{\textrm{c},1}}}$}
 \multiput(0,0)(0,5){5}{
    \multiput(0,0)(5,0){4}{\put(0,0){${-}$}}
    }
 \multiput(0,5)(0,10){2}{
    \multiput(0,0)(10,0){2}{\put(0,0){${+}$}}
    }
 \multiput(5,0)(0,10){3}{
    \multiput(0,0)(10,0){2}{\put(0,0){${+}$}}
    }
 \put(20,15){${+}$}
 \put(30,16){\vector(0,1){6}}
 \put(30,6){\vector(0,-1){6}}
 \put(28,9){${\ell_{\textrm{c}}=\lfloor\frac{2}{h}\rfloor+1}$}
 \end{picture}
\caption{Graphical description of $\Gamma_\textrm{m}$ for the 
nearest neighbor PCA.}
\label{f:drop-nn}
\end{figure}
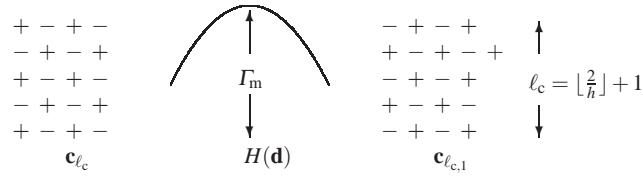

In the nearest neighbor case ($\kappa=0$) it has been proven \cite{CN}
that the set of metastable states 
is $\spazio_\textrm{m}=\{\muno,\chess_\textrm{e},\chess_\textrm{o}\}$. 
It is important to note that the two states 
$\chess_\textrm{e}$ and $\chess_\textrm{o}$ are essentially the same 
metastable state, indeed, it can be easily seen that 
$\chess_\textrm{e}$ is the downhill image of $\chess_\textrm{o}$ and 
vice-versa. So that, when the system is trapped in such a metastable 
state, it flip--flops between these two configurations. 
Moreover, it has also been proven that the maximal stability level 
is given by 
\begin{equation}
\label{msl-nn}
\Gamma_\textrm{m}
=H_h(\textbf{c}_{\ell_{\textrm{c}}})
   +\Delta_h(\textbf{c}_{\ell_{\textrm{c}}},
             \textbf{c}_{\ell_{\textrm{c},1}})-H_h(\muno)
\stackrel{\beta\to\infty}{\sim}
\frac{8}{h}
\end{equation}
where $\ell_\textrm{c}=\lfloor 2/h\rfloor+1$ is called 
\textit{critical length}, 
$\textbf{c}_{\ell_{\textrm{c}}}$ is a configuration characterized 
by a $\ell_\textrm{c}\times(\ell_\textrm{c}-1)$ 
rectangular checkerboard droplet in the sea of minuses, and 
$\textbf{p}_{\ell_{\textrm{c},1}}$ is a configuration characterized 
by a $\ell_\textrm{c}\times(\ell_\textrm{c}-1)$ 
rectangular checkerboard droplet in the sea of minuses
with a single site plus protuberance attached to one of the two longest 
sides of the rectangle (see figure~\ref{f:drop-nn}).
It is worth noting that, compare \eqref{msl-cr} and \eqref{msl-nn}, 
the exit from the metastable state is much slower in the case of the 
cross PCA with respect to the nearest neighbor one. 

Even in this case the properties of the metastable 
states are an immediate consequence of the Theorem stated 
above. But also for the nearest neighbor PCA the nucleation 
property is proven: during the transition 
during the escape from the 
metastable state $\muno$ to the stable one $\puno$ 
the chain visits with probability tending to one in the limit 
$\beta\to\infty$ the configuration 
$\textbf{c}_{\ell_{\textrm{c}}}$ and, starting from such a 
configuration, it performs the jump to 
$\textbf{c}_{\ell_{\textrm{c},1}}$.
From the physical point of view this property means that the escape 
from the metastable state is achieved via the nucleation of the 
critical checkerboard droplet $\textbf{c}_{\ell_{\textrm{c}}}$.

Moreover, in the nearest neighbor case it has been proven that 
during the escape from $\muno$ to $\puno$ the system 
has also to visit the checkerboard metastable states 
$\{\chess_\textrm{e},\chess_\textrm{o}\}$. Starting from such 
a metastable state, the system performs the final escape to 
$\puno$ with an exit time controlled by the same maximal stability 
level $\Gamma_\textrm{m}$ \eqref{msl-nn}. 

Finally, we just mention the heuristic results discussed in \cite{CNS2}
for the tuned cross PCA with $0<\kappa<1$. There is one single 
metastable state, i.e., $\spazio_\textrm{m}=\{\muno\}$, but, 
depending on the ration $\kappa/h$, the system exhibits 
different escaping mechanisms. 
In particular, for $h<2\kappa$ the systems perform a direct 
transition from $\muno$ to $\puno$, whereas for $2\kappa<h$
the system ``necessarily" visits the not metastable 
checkerboard state before touching $\puno$.
In \cite{CNS2} it has been pointed out the analogies between 
the behavior of the tuned cross PCA and the Blume--Capel model 
\cite{CO}. The metastable character of the two models 
is very similar with the role of the self--interaction parameter $\kappa$
played by that of the chemical potential in the Blume--Capel model.


%
%
%

\end{document}